\documentclass[epj]{svjour}

\usepackage{amsmath}
\usepackage{graphicx}
\usepackage{fancyhdr}

\def\mytitle{My title} 
\def\myauthors{My name}  
\def\mytype{My type of session}
\def\mysession{My session}
\def\mytitle{Time-dependent CP asymmetries in $\mathrm{B}_s$ decays at LHCb} %Put your title here!
\def\myauthors{Johan Blouw on behalve of the LHCb collaboration}    %Put your name here!
\def\mytype{Contributed Talk}    
\def\mysession{Flavor Physics}
\begin{document}
%
% Definitions:
%
\newcommand{\bs}{\mathrm{B}_s}
\newcommand{\bsbar}{\bar{\mathrm{B}}_s}
\newcommand{\ds}{\mathrm{D}_s}
\newcommand{\jpsi}{J\!/ \Psi}
\newcommand{\phiphi}{\phi \phi}
\newcommand{\jpsism}{\jpsi \phi}
\newcommand{\bstojpsiphi}{\bs \to \jpsism}
\newcommand{\bsmjpsiphi}{\bs \to \jpsism}
\newcommand{\bsbmjpsiphi}{\bar{\mathrm{B}}_s \to \jpsism}
\newcommand{\bsdspi}{\bs \to \ds \pi}
\newcommand{\bstophiphi}{\bs \to \phiphi}
\title{Time-dependent CP asymmetries in $\mathrm{B}_s$ decays at LHCb}
\author{Johan Blouw, on behalf of the LHCb collaboration.
\thanks{\emph{Email:}johan.blouw@physi.uni-heidelberg.de}%
}                     % Do not remove
\institute{Physikalisches Institut,\\
Philosphenweg 12\\
69120 Heidelberg, Germany
}
\date{}
\abstract{
The LHCb experiment will search for New Physics in $\bs$  mixing.
The $\bs$ mixing phase will be extracted from the measurement of  
the time-dependent CP asymmetry in exclusive $\mathrm{B}_s$ decays governed by  
the $b \rightarrow c\bar{c} s$ quark-level transition. 
Large New Physics effects can  be discovered or excluded with the 
data collected during the very first physics run of LHC. 
Based on Monte Carlo simulations of the LHCb detector, the expected
sensitivity with $2~\mathrm{fb}^{-1}$ on the CP-violation parameter
$\phi_s$, is $\sigma(\phi_s) = 0.022$.
%The latest expectations, based on Monte Carlo  studies with fully simulated 
%events, will be presented. The LHCb  sensitivity to other CP-violating 
%observables in $\bs$ decays will be  reviewed.
\PACS{
	{14.40.Nd}{Bottom mesons} \and
	{13.25.Hw}{Decays of bottom mesons} \and
	{11.30.Er}{Charge conjugation, parity, time reversal, and other discrete symmetries} 
}
} %end of abstract
\maketitle
\section{Introduction}
\label{intro}
In the Standard Model (SM), CP-violation in flavour changing currents 
is caused by one single phase in the mixing matrix. This matrix describes
the charged current weak interactions of quarks~\cite{rf:nir_1}.
This so-called Cabbibo-Kobayashi-Maskawa (CKM) matrix is a complex, unitary, $3 \times 3$ matrix
which relates the electroweak eigen-states of the down-type quarks with their mass eigen-states.
From the unitarity requirement of the CKM matrix 6 orthogonality relations can be derived,
which are usually displayed as unitarity triangles in a complex plane.

$\mathrm{B}$-meson decays allow to test some of these relations.
An example is the unitarity relation 
\begin{equation}\label{eq:ckm}
\mathrm{V}_\mathrm{ub}^* \mathrm{V}_\mathrm{us} + \mathrm{V}_\mathrm{cb}^* \mathrm{V}_\mathrm{cs}
+ \mathrm{V}_\mathrm{tb}^* \mathrm{V}_\mathrm{ts} = 0.
\end{equation}
which can be studied using $b \rightarrow c\bar{c} s$ quark-level transitions of the $\mathrm{B}_s$ meson.
The elements $\mathrm{V}_{xy}$ describe the complex coupling strengths of the up-type quarks $(x = u,c,t)$ to
the down-type quarks, $(y = d,s,b)$. When using the Wolfenstein parameterisation of the CKM
matrix~\cite{rf:wolfenstein}, it can be shown that the three terms of Eq.~\ref{eq:ckm} relate to each other as 
$\mathcal{O}(\lambda^4)~:~\mathcal{O}(\lambda^2)~:~\mathcal{O}(\lambda^2)$. Here, $\lambda = sin(\theta_C)$ is the
expansion parameter, with $\theta_C$ being the Cabbibo angle.
The angle between the two larger sides $(\mathrm{V}_\mathrm{cb}^* \mathrm{V}_\mathrm{cs})$
and $(\mathrm{V}_\mathrm{tb}^* \mathrm{V}_\mathrm{ts})$ is small:
\begin{equation}
\chi = \arg \left[ -\frac{\mathrm{V}_\mathrm{cb}^* \mathrm{V}_\mathrm{cs}}
	{\mathrm{V}_\mathrm{tb}^* \mathrm{V}_\mathrm{ts}} \right] 
	\approx \lambda^2 \eta \approx \arg(\mathrm{V}_\mathrm{ts}) - \pi \approx 0.02
\end{equation}

The neutral $\mathrm{B}_s$ meson undergoes mixing which can be described by
an effective Hamiltonian consisting of two $2 \times 2$ matrices: the mass matrix $M$, 
and decay matrix $\Gamma$. The mass difference between the mass eigen
states is defined as $\Delta m_s = m_H - m_L$, and the corresponding decay-time 
difference by $\Delta \Gamma_s = \Gamma_L - \Gamma_H$, where $(L,H)$ 
indicate the heavy and light mass state.
Since the mass eigen-states are not equal to the 
weak eigen-states, these matrices contain off-diagonal elements, $M_{12}$ and $\Gamma_{12}$. 
%Transitions between the weak eigen-states are caused by the off-diagonal matrix elements.
The phase difference between $M_{12}$ and $\Gamma_{12}$ leads to an observable CP-violating phase, 
$\phi_s$. In the SM it is related to $\chi_s$ through 
$\phi_s \approx 2 \arg(\mathrm{V}_\mathrm{tb}^*\mathrm{V}_\mathrm{ts})$.

Experimentally, the angle $\phi_s$ is measured by evaluating the time-dependent CP asymmetry
of the $\bs \to \jpsi \phi$ decay,
\begin{equation}
\mathcal{A}_\mathrm{CP}(t) = \frac{\Gamma(\bsbar(t) \to f) - \Gamma(\bs(t) \to f)}{\Gamma(\bsbar(t) \to f) + \Gamma(\bs(t) \to f)}
\end{equation}
which is constructed from the time-dependent decay rates $\Gamma$ for initial $\bsbar$ mesons and $\bs$ mesons
to the same finale state $f = \jpsi \phi$.
This can then be parameterised as
\begin{equation}\label{eq:cp_osci}
\mathrm{A}_\mathrm{CP}(t) = \frac{-\eta_f \sin\phi_s \sin(\Delta m_s t)}
	{\cosh(\Delta\Gamma_s \frac{t}{2}) - \eta_f \cos\phi_s\sinh(\Delta\Gamma_s \frac{t}{2})},
\end{equation}
%\begin{equation}
%\mathcal{A}_\mathrm{CP}(t) = -\frac{\mathcal{A}_\mathrm{CP}^\mathrm{dir}\cos(\Delta M_s t) + \mathcal{A}_\mathrm{CP}^\mathrm{mix-ind}\sin(\Delta M_s t)}
%	{\cosh(\Delta\Gamma_s t/2) + \mathcal{A}_{\Delta \Gamma_s}\sinh(\Delta \Gamma_s t/2)},
%\end{equation}
where $\eta_f = +1,-1$ for the CP-even, CP-odd sub sample respectively.
The mass-difference between the mass eigen-states of the 
$\mathrm{B}^0_s$ meson determines the oscillation frequency of this asymmetry.

Although the SM predictions of CP-violating observables are consistent with all measurements
so far, there is also a clear need for physics beyond the SM because 
it is unable to account for the observed baryon asymmetry in 
the universe~\cite{rf:dolgov_1,rf:dolgov_2}.
%since the SM is unable to account for the
%baryon density number $\mathrm{Y}_B = (n_B - n_{\bar{B}})/s \approx 9 \times 10^{-11}$.
%Here, $s$ is the entropy density of the universe at a certain temperature. 
%For a more detailed discussion of these estimates, see \emph{e.g.} 
%References~\cite{rf:dolgov_1,rf:dolgov_2}.
%%This leads to the conclusion that there must exist CP-violation beyond the SM.
%In addition, there is no explanation why CP is so small in QCD. This so-called 'strong CP problem'
%also might be a clue to New Physics~\cite{rf:nir_1}.
%Most extensions of the SM provide possible new sources of CP-violation. 
Most Extension of the SM may lead to new CP-violating phases, and therefore 
to non-SM contributions to CP-observables~\cite{rf:nir_1}.
%If there is new physics below the
%$1~\mathrm{TeV}$ mass scale, there may be new CP-violating phases, and there may
%be non-SM contributions to CP-observables

\section{The LHCb spectrometer}
\begin{figure*}[t]
\label{fig:lhcb}
\includegraphics[width=1.0\textwidth]{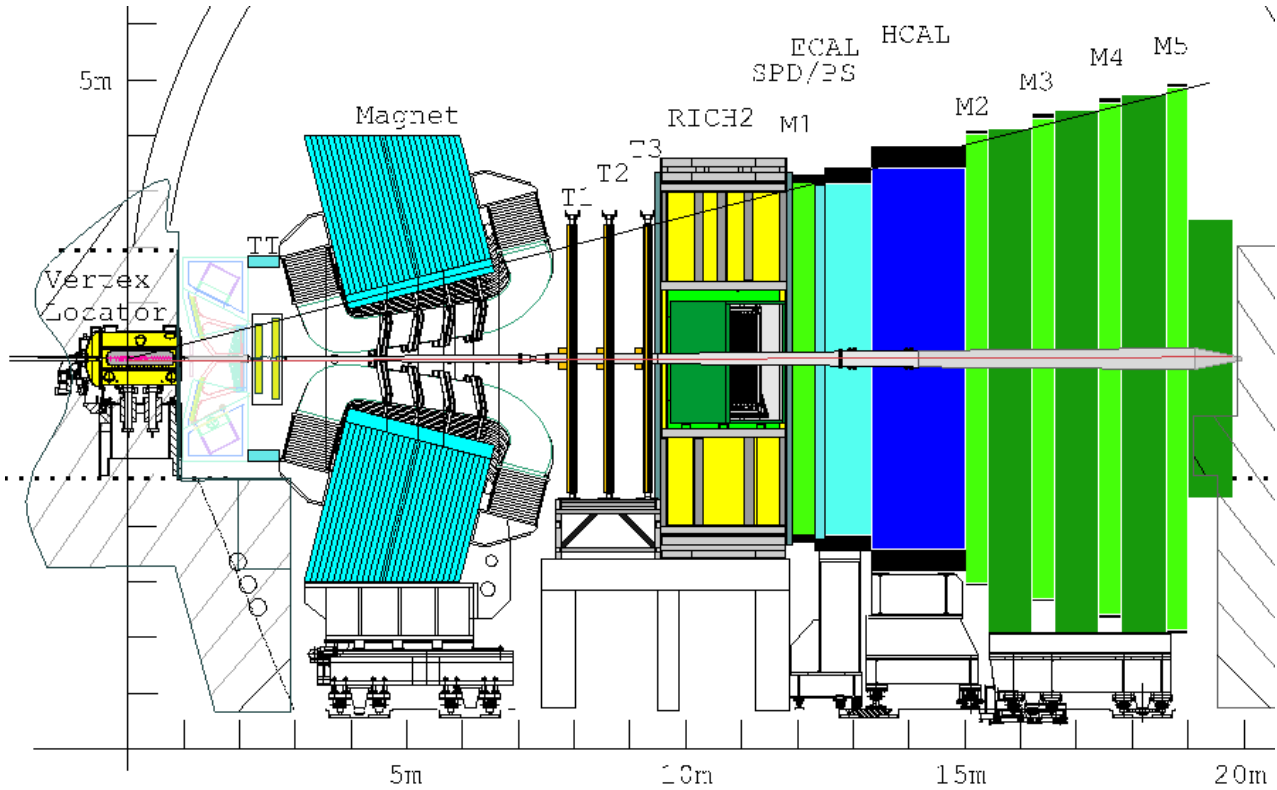}
\caption{The LHCb spectrometer, from left to right: the Vertex Locator (VeLo), positioned at the
interaction point, the RICH 1 particle identification detector, the TT which is a silicon strip
detector, the $4.2~\mathrm{Tm}$ dipole magnet, the Outer Tracker and Inner Tracker, RICH 2,
the electromagnetic and hadronic calorimeters and finally the Muon system.}
\end{figure*}

The LHCb spectrometer is a dedicated B-physics experiment for the Large Hadron Collider, at CERN 
and is currently under construction.
Protons from both beams collide, and subsequently $b\bar{b}$ quark pairs are produced with a 
cross section of about $500~\mu\mathrm{b}$. 
Due to the fact the $b\bar{b}$ pairs are produced
with a large boost, a forward spectrometer can reconstruct the decay products of the
%in the forward direction,
two B-mesons. Figure~\ref{fig:lhcb} shows a schematic of the LHCb spectrometer; the
VErtex detector (VELO) is located around
the interaction point, a second silicon tracker (TT) in front of the $4.2~\mathrm{Tm}$ dipole magnet
and together with the Inner Tracker (IT) and Outer Tracker (OT) behind the magnet, comprise the 
tracking system. Between VELO and TT, a Ring Imaging CHerenkov (RICH) is situated for particle identification.
Behind the IT $\&$ OT, there is a second RICH detector, followed by an electromagnetic
and a hadronic calorimeter. Muons are detected by the Muon System positioned behind the calorimeters.
Monte Carlo studies show that a tracking efficiency better than $95\%$ and a momentum resolution
between $0.4\%$ and $0.6\%$ is feasible. 
%The spatial resolution of the reconstructed $\mathrm{B}_s$
%depends on the decay and in the resolution along the z-direction. For the $\bs \to \jpsi \phi$, this
%is Proper time:

\section{New Physics from $\bs$ decays}
The $\bs \to \jpsi \phi$ decay can proceed through a tree diagram, 
where a $b \to c$-quark transition is mediated by a W-boson.
The final state, $\jpsi \phi$, can also be reached through a box diagram where
the $\bs$-meson first oscillates into a $\bar{\mathrm{B}}_s$ meson, before
it decays. 
In the $\bs \to \phi \phi$ case, the leading amplitude is a penguin diagram.
Similar to the previous case, the $\bs$ meson can oscillate before it decays.

\begin{figure}[t]
\includegraphics[width=0.45\textwidth]{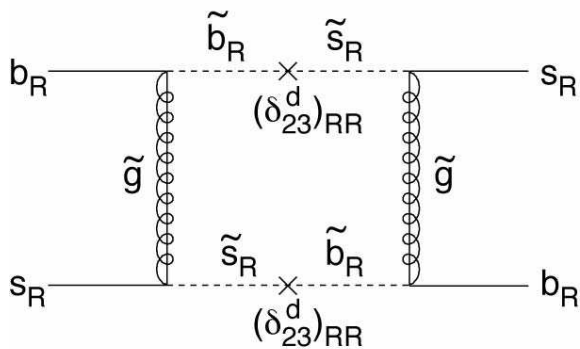}
\caption{Possible SUSY contribution (gluino exchange) to $\bs - \bar{\bs}$ mixing.}
%\caption{Example of a Feynman box diagram where gluinos are exchanged.}
\label{fig:gluino_box}       % Give a unique label
\end{figure}

\begin{figure}[hbt]
\includegraphics[width=0.45\textwidth]{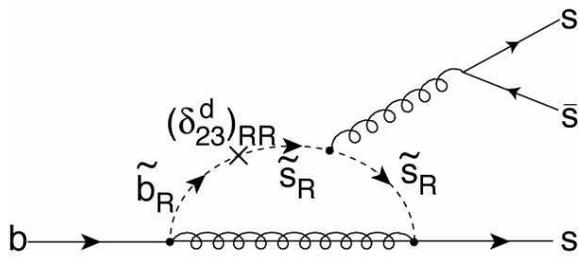}
\caption{Example of a penguin diagram with SUSY contributions in the form of gluino exchange.}
\label{fig:gluino_exchange}       % Give a unique label
\end{figure}

New Physics (NP) can be observed in B-meson decays, through contributions from loop-diagrams. 
Here, the exchange of new particles in the box-diagram describing $\bs - \bar{\bs}$ mixing 
of Fig.~\ref{fig:gluino_box} could lead to new non-SM phases, and therefore observable deviations
from SM CP-violation. In a similar way, new particles can contribute
to a b-quark decay in the loop of a penguin diagram, see Fig.~\ref{fig:gluino_exchange}.
In this article the following parameterisation of NP is used for the 
off-diagonal elements of the mass-mixing matrix, 
\begin{equation}
M^\prime_{12} = M_{12}(1 + h_s e^{2i\sigma_s}).
\end{equation}
Here, $h_s$ denotes the effective scale of the NP contribution with an effective 
phase $\sigma_s$. The afore mentioned decay, $\bs \to \jpsi \phi$ is sensitive 
to the additional contribution
to the mixing. The Standard Model predicts the CP-asymmetry 
$\mathcal{A}_\mathrm{CP} \sim \sin\phi_s$ 
in this channel to be very small: 
\begin{equation}
\phi_s(\bstojpsiphi) \approx -2\chi = -0.035~\mathrm{rad}.
\end{equation}
%hence this channel could be sensitive to contributions from NP.

For the decay $\bstophiphi$, the Standard Model does not predict any CP-asymmetry
%\begin{equation}
%\phi_s^\mathrm{SM} \approx 2 \arg(V_{\mathrm{ts}}^* V_\mathrm{tb}) - \arg(\frac{V_\mathrm{tb}V_\mathrm{ts}^*}{V_\mathrm{tb}^* V_\mathrm{ts}}) 
%= -2\chi + 2 \chi = 0,
%\end{equation}
and therefore any observation of CP-violation is a clear indication of New Physics.
%This can be quantified using the previously mentioned parameterisation. In the $\bsmjpsiphi$ case,
%the weak mixing angle $\phi_s^\mathrm{SM}$ receives a contribution of $-\arg[1 + h_s e^{2i\sigma_s}]$ from possible
%NP. 
%%Note that $\Delta m_s^\mathrm{SM}$ and $\Delta\Gamma_s^\mathrm{SM}$ also receive contributions from NP.

%For a measurement of time-dependent CP asymmetries, the CP odd/even components of the final state have 
%to be extracted. In some decays the final state is always
%CP-even, \emph{e.g.} $\bs \to \jpsi \eta$, $\bs \to \eta_c \phi$, $\bs \to \mathrm{D}_s^+\mathrm{D}_s^-$, and $\bsdspi$.
%In the case of $\bstojpsiphi$ and $\bstophiphi$, the final state is an admixture of CP-odd and CP-even components.
%In the latter decays, an angular analysis of the final-state components allows for the separation of CP-even
%and CP-odd components. 

\section{LHCb sensitivity to mixing phase $\phi_s$}
The evaluation of the LHCb sensitivity to $\phi_s$ proceeds in two steps.
First a detailed Monte Carlo simulation of physics and LCHb detector
response is used to determine signal and background yields, efficiencies,
and resolutions. These results have been used in fast MC simulations to 
determine the sensitivity to $\phi_s$ from different decay channels.

\begin{figure}[th]
\includegraphics[width=0.45\textwidth]{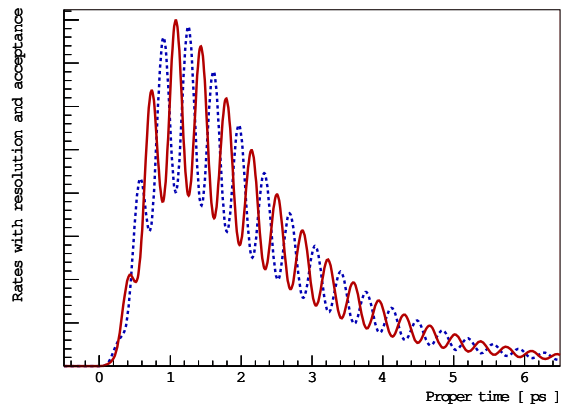}
\caption{$\mathrm{B}_s \to \mathrm{D}_s^- \pi^+$ decay rates \emph{vs.} 
proper $\bs$ decay time.
The solid line shows the oscillation of an initially produced $\bs$ meson, 
the dashed line is for an initially $\bar{\mathrm{B}}_s$ meson.}
%%\caption{The oscillations observed in case of perfect reconstruction are diminished by 
%%the effect of flavour tagging (blue curve), proper=time resolution (yellow curve), 
%%background (purple) and finally, largely reduced by the acceptance of the LHCb spectrometer.}
\label{fig:oscillations}       % Give a unique label
\end{figure}

As can be seen from Eq.~\ref{eq:cp_osci}, the observed time-dependent 
CP-asymmetry is modulated by the oscillations from mixing.
The oscillation frequency can be best measured using the $\bsdspi$ decay. 
The observed time-dependent decay rate is given by
\begin{equation}
R(t) \propto \frac{e^{-\Gamma_s t}}{2} 
\left\{ \cosh\frac{\Delta\Gamma_st}{2} + rD\cos(\Delta m_s t) \right\}.
\end{equation}
Here, $r = +1$ for $\bs$ mesons and $r = -1$ for $\bar{\bs}$ mesons.
The dilution $\mathrm{D}$ is given by the probability that a $\bs$ meson was wrongly tagged.
In one year of LHCb running ($2~\mathrm{fb}^{-1}$), a yield of 140k events is 
expected, with a $B/S < 0.05$ at a tagging efficiency of $\epsilon_\mathrm{tag} \approx 0.6$ 
and $D \approx 0.4$. The rate as a function 
of proper time is shown in Figure~\ref{fig:oscillations}. From this a
statistical accuracy of $\sigma(\Delta m_s) = 0.006~\mathrm{ps}^{-1}$ is estimated. 
Systematic uncertainties have not been considered.

\begin{figure}[b]
\includegraphics[width=0.45\textwidth]{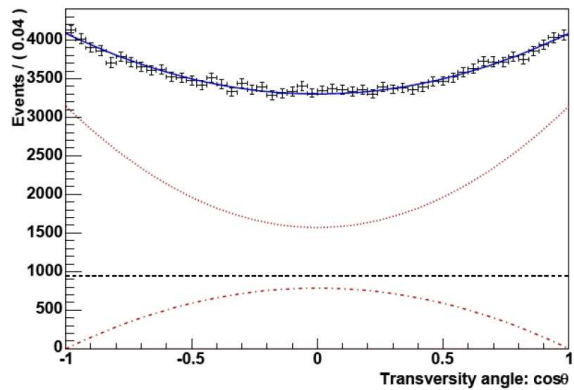}
\caption{Transversity angle distribution of the $\bstojpsiphi$ decay. The solid line through the data-points
represents the sum of the contributions from the CP-even (dotted line), the CP-odd (dot-dashed) components and from background (dashed).}
%\caption{Schematic diagram of the $\jpsi - \phi$ system in the centre of mass of the $\jpsi$. In that system,
%$\theta_\phi$ is the opening angle between the two Kaons, and they span the plane in which the polar 
%angle $\Theta_\mathrm{tr}$ between the positive $\mu$ and the positive $z$-axis is called 
%the transversity angle, $\phi_\mathrm{tr}$ is the corresponding azimuthal angle.}
\label{fig:transversity}       % Give a unique label
\end{figure}

The CP-odd and even components are separated from each other on a 
statistical basis by employing a transversity-angle analysis. 
The angular dependence of the decay-rate is given by
\begin{equation}
\frac{d\Gamma}{dc} \propto \left[ \left| \mathrm{A}_0 \right|^2 + \left|\mathrm{A}_\parallel \right|^2 \right] 
	\frac{3}{8} (1+c^2) + \left| \mathrm{A}_\perp 
	\right|^2\frac{3}{4}(1-c^2),
\end{equation} 
with  $c = \cos \Theta_{\mathrm{tr}}$; $\mathrm{A}_\parallel$ and 
$\mathrm{A}_0$ are the CP-even components, and $\mathrm{A}_\perp$ 
is the CP-odd component. The transversity angle $\Theta_\mathrm{tr}$ is 
defined in the rest-frame of the $\jpsi$ as the angle between
the positive muon and the z-axis, which is perpendicular to the plane 
spanned by the two decay products of the $\phi$. The transversity angle 
distribution is shown in Figure~\ref{fig:transversity}.
It can be described by the sum of the CP-even, CP-odd and background components.

%In the case of the $\bs \to \phi \phi$ decay, care is taken of the fact that the 2 Kaon pairs observe Bose statistics,
%and hence are treated symmetrically. 
%For the $\bs \to \phi \phi$ decay, in $2~\mathrm{fb}^{-1}$ a yield of 3.1k events is expected, with a $B/S < 0.8$,
%yielding a statistical accuracy of $\sigma(\phi_s) = 0.11$.

In addition to the parameters which describe detector effects, (see above)
the following physics parameters were used as input to the fast simulation:
$\phi_s$, $\Delta m_s$, $\Delta\Gamma/\Gamma$, lifetime of the $\bs$ ($\tau_{\bs})$ and $R_T$, the observable
fraction of CP-odd eigenstates, as shown in Table~\ref{tab:phys_inp},
\begin{table}
\caption{Input parameters to the fast simulation.}
\label{tab:phys_inp}
\begin{tabular}{|c|c|c|c|c|}
$\phi_s [\mathrm{rad}]$ & $\Delta\mathrm{M}_s [\mathrm{ps}^{-1}]$ & $\Delta\Gamma_s/ \Gamma_s$ & $\tau_{\mathrm{B}_s^0}[\mathrm{ps}]$ & $R_T$ \\\hline
-0.04 & 17.5 & 0.15 & 1.45 & 0.2
\end{tabular}
\end{table}
where these values have been taken from literature (\emph{e.g.} Ref.~\cite{rf:pdg2004}). 
In addition to the mixing phase $\phi_s$, the channel $\bs \to \jpsi \phi$ also allows
for the determination of the width difference $\Delta \Gamma_s$. The predicted sensitivity 
for $\Delta \Gamma_s / \Gamma_s$ is $\sigma(\Delta \Gamma_s / \Gamma_s) = 0.008$.

%\begin{table}[htb]
%\caption{Results from full MC studies for the sensitivity to
%the relative life-time differences for CP-even and the CP-mixed channels, respectively.
%The statistical errors are quoted for an integrated luminosity of $2~\mathrm{fb}^{-1}$.}
%\label{tab:sens_mass}
%\begin{tabular}{c|c|c|c|c|c}
%Sensitivity & $\jpsi \eta$      & $\jpsi  \eta $ & $\eta_c \phi$ & $\mathrm{D}_s^+\mathrm{D}_s^-$ & $\jpsism$ \\
%            & $(\gamma \gamma)$ & $(3\pi)$       &               &                                &          \\\hline
%%$\sigma(\mathrm{B}_s)[\frac{\mathrm{MeV}}{\mathrm{c}}]$ & 34 & 20 & 12 & 6 & 14 \\
%$\sigma(\Delta \Gamma_s/ \Gamma_s)$ & 0.016 & 0.019 & 0.018 & 0.017 & 0.0079 \\
%\end{tabular}
%\end{table}

Finally, the predicted uncertainties on the measurement of the weak phase $\phi_s$ 
are given in Table~\ref{tab:pred_phis}.
Here, the sensitivity extracted from the five $b \to c\bar{c}s$ decays 
is predicted to be $\sigma(\phi_s) = 0.022$, where the $\bs \to \jpsi \phi$ is the most 
sensitive channel.
\begin{table}[hb]
\caption{Summary of the estimated LHCb sensitivity for a data set of $2~\mathrm{fb}^{-1}$.
The relative weight of the five channels to the determination of $\phi_s$ is given as well.
%the contributions of CP-even final states $(\jpsi \! \eta(3\pi),~\jpsi \! \eta(\gamma \gamma),
%~\mathrm{D}_s \mathrm{D}_s~\mathrm{and}~\eta_c \phi)$, and the final state consisting of an admixture
%of CP-even and CP-odd states, $\jpsi \phi$.
}
\label{tab:pred_phis}
\begin{tabular}{l|c|c}
%%Channel & $\sigma(\phi_s)[\mathrm{rad}]$ & Weight $(\sigma/\sigma_i)^2~[\%]$ \\\hline
Channel & $\sigma(\phi_s)[\mathrm{rad}]$ & Weight$[\%]$ \\\hline
$\mathrm{B}_s \to \jpsi \! \eta(3\pi)$ & 0.14 & 2.3\\
$\mathrm{B}_s \to \mathrm{D}_s^+ \mathrm{D}_s^-$ & 0.13  & 2.6 \\
$\mathrm{B}_s \to \jpsi \! \eta(\gamma \gamma)$ & 0.11 & 3.9 \\
$\mathrm{B}_s \to \eta_c \phi$ & 0.11 & 3.9 \\ \hline
Combined sensitivity : & 0.06 & 12.7 \\ \hline
$\bsmjpsiphi$ & 0.023 & 87.3 \\ \hline
Total combined sensitivity: & 0.022 & 100.0 \\ \hline
\end{tabular}
\end{table}

\begin{figure}[htb]
% For example, with the option graphicx use
\includegraphics[width=0.45\textwidth]{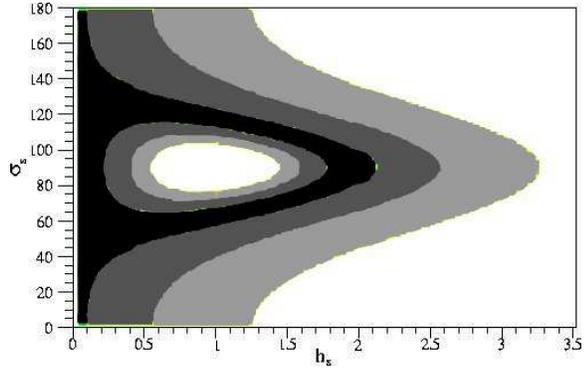}
\caption{Limits on the NP parameters $\sigma_s$ and $h_s$ taken from  Ref.~\cite{rf:ligeti_1}.
The shaded areas correspond to allowed parameter values 
with confidence level $\mathrm{CL} > 0.9$ (black), $\mathrm{CL} > 0.32$ (dark grey) and
$\mathrm{CL} > 0.05$ (light grey), respectively. The white area around $h_s = 1, \sigma_s = 90$ is caused by cancelling
contributions to $\Delta m_s$~\cite{rf:ligeti_1}.}
\label{fig:before}       % Give a unique label
\end{figure}

\begin{figure}[htb]
% For example, with the option graphicx use
\includegraphics[width=0.45\textwidth]{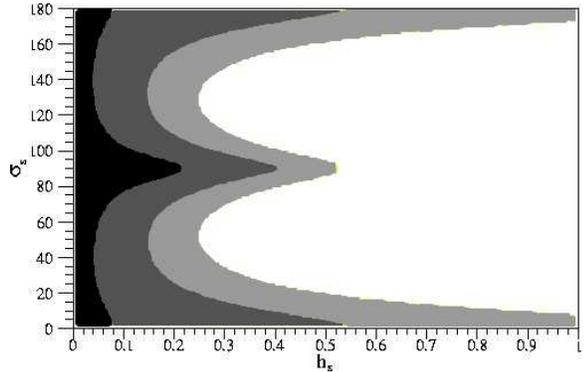}
\caption{Limits on $\sigma_s$ and $h_s$ from Ref.~\cite{rf:ligeti_1} using MC predictions for
LHCb measurements of $\phi_s$, $\Delta m_s$ and $\Delta \Gamma_s/ \Gamma_s$.
The shaded areas show allowed regions for $\sigma_s$ and $h_s$ for $CL > 0.9$ (black), $\mathrm{CL} > 0.32$ (dark grey) and $\mathrm{CL} > 0.05$ (light grey).
}
\label{fig:after}       % Give a unique label
\end{figure}

These results can then be used to predict limits on NP by using the previously mentioned parameterisation.
In an article by Ligeti~\cite{rf:ligeti_1}, the reported LHCb sensitivities 
have been used to estimate limits on $\sigma_s$ and $h_s$. 
From Ref.~\cite{rf:ligeti_1}, two figures have been taken showing exclusion limits 
on $\sigma_s$ and $h_s$. These limits have been obtained including the 
first measurements of $\Delta m_s$~\cite{rf:cdf} (Figure~\ref{fig:before}).
With the predicted LHCb accuracies on $\phi_s$ and $\Delta \Gamma_s/\Gamma_s$ as reported here, 
the exclusion limits on the amplitude $h_s$ are significantly improved see Figure~\ref{fig:after}.

\section{Conclusions}
Despite the success of the Standard Model in predicting and explaining
its parameters as measured by experiments, 
%there are some clear discrepancies in the CP-violation sector. 
%In particular, CP-violation in the $\mathrm{B}_s$-meson system is
NP leading to additional CP-violation is necessary to explain the baryon asymmetry in the universe.
CP-asymmetries of $\mathrm{B}_s$-meson decays are expected to be sensitive probes of
New Physics. The SM predicts small asymmetries for the $b \to c \bar{c}s$ transitions of $\bs$-meson,
such as $\bs \to \jpsi \phi$. Through the contribution of phase from NP, significantly
larger CP-violation may occur. The LHCb collaboration is currently
constructing a dedicated B-physics spectrometer for the LHC to provide
accurate measurements of CP-violation in many rare $\mathrm{B}$ decay channels.
From Monte Carlo studies of the channels 
$\bs \to \jpsi \phi$, $\bs \to \mathrm{D}_s^+ \mathrm{D}_s^-$, $\bs \to \jpsi \eta(\gamma \gamma)$,
$\bs \to \jpsi \eta(3\pi)$ and $\bs \to \eta_c \phi$ and $\bs \to \mathrm{D}_s^- \pi^+$,
a sensitivity for the CP-violating phase $\phi_s$, of $\sigma(\phi_s) = 0.022$ is found.
%The result is a possible improvement of a factor 10 in the limits set on contributions
%from NP.

\end{document}